\documentclass{article}
\usepackage{array}
\usepackage[utf8]{inputenc}
\usepackage{amsmath}
\usepackage{amsthm}
\usepackage{amssymb}
\usepackage{mathrsfs}
\usepackage{bbm}
\usepackage{hyperref}
\usepackage[authoryear]{natbib}
\usepackage[margin=1.25in]{geometry}
\usepackage{setspace}
\usepackage[usenames,dvipsnames]{xcolor}
\usepackage{ulem}%for underlines
\usepackage{authblk}
\usepackage{fancyhdr} %for headers
\usepackage{graphicx}
\usepackage{tikz}
\usepackage{comment}
\usetikzlibrary{patterns}
\usepackage{caption}
\usepackage{subcaption}
\usepackage{dsfont}

\hypersetup{colorlinks, citecolor = Blue, linkcolor = Green, urlcolor = MidnightBlue}

\title{The Effect of the U.S. Federal Production Tax Credit on Negative Energy Prices}
\author[1]{Erik P. Johnson}
\author[2]{Daniel H. Karney}
\affil[1]{Carthage College\footnote{Corresponding author. epj@carthage.edu, 2001 Alford Park Dr., Kenosha, WI 53140, USA.}}
\affil[2]{Ohio University\footnote{karney@ohio.edu, 1 President St., Athens, OH 45701, USA.}}
\date{\today}

\begin{document}

%%%%%%%%%%%%%%%%%%%%%%%%%%%%%%%%%%%%%%%%%%%%%%%%%%

\maketitle
\thispagestyle{empty}

%%%%%%%%%%%%%%%%%%%%%%%%%%%%%%%%%%%%%%%%%%%%%%%%%%
%%%%%%%%%%%%%%%%%%%%%%%%%%%%%%%%%%%%%%%%%%%%%%%%%%

\begin{abstract}
\normalsize
\noindent Negative prices frequently arise in electricity markets and production subsidies are commonly cited as a leading cause. Using data from the U.S. Eastern Interconnection wholesale electricity markets for the period 2000-2023, we document that at least 18 percent of the total hours with negative energy prices cannot solely be attributed to the federal production tax credit (PTC), a large, persistent production subsidy during the period of study. Next, we provide the first causal evidence that the PTC leads to negative energy prices. On average, increasing PTC-eligible wind generation by 1 percent increases the probability of a negative energy price by 3.4 percentage points. However, significant heterogeneity exists across electricity markets, and some markets show an effect that is an order of magnitude larger. Furthermore, the PTC explains only 10 percent of the observed negative prices that could potentially be attributable to it. Finally, we use the causal estimate to calculate a plausible value for the environmental benefits accrued from the PTC subsidy when energy prices are negative.
\end{abstract}
\vspace{2mm}



\vspace{1cm}

\noindent We are grateful for comments and suggestions from Dylan Brewer and Benjamin Edelstein as well as participants at the 2025 USAEE/IAEE North American Conference and 2026 AERE@MEA. All mistakes are our own. No AI was used in writing this manuscript, but was employed in editing STATA code.

%%%%%%%%%%%%%%%%%%%%%%%%%%%%%%%%%%%%%%%%%%%%%%%%%%
%%%%%%%%%%%%%%%%%%%%%%%%%%%%%%%%%%%%%%%%%%%%%%%%%%
%%%%%%%%%%%%%%%%%%%%%%%%%%%%%%%%%%%%%%%%%%%%%%%%%%

\doublespacing
\newpage
\setcounter{page}{1}

\noindent Negative prices are an uncommon phenomenon in most areas of the economy, but surprisingly common in wholesale electricity markets. \cite{seel2021plentiful} documents that negative prices occurred in four percent of all hours in the seven organized U.S. wholesale electricity markets in 2020.\footnote{Negative prices also occur in electricity markets outside the United States. See \cite{schneider2010power}, \cite{keles2012comparison}, \cite{fanone2013case}, \cite{brijs2015statistical}, \cite{de2015negative}, \cite{gerster2016negative}, \cite{aust2020negative}, \cite{frondel2022market}, and \cite{prokhorov2022impact} for studies of negative prices in the European context.} Negative electricity prices reverse the usual buyer-seller relationship in the short run as well as increase the risk of investment in new generation capacity in the long run \citep{huntowski2012negative}. Additionally, as \cite{sher2013wind} describes, this role reversal can reduce the profitability of base load capacity, such as coal- and gas-fired power plants, accelerating retirements and leading to reduced grid reliability and more volatile prices. In our data set covering the years 2000-2023 for all electricity markets in the U.S. Eastern Interconnection, payments by electricity suppliers totaled \$4.5 billion, thus indicating the scale of the negative prices phenomenon.

Production subsidies for renewable generation – specifically, the federal production tax credit (PTC) for wind and solar generation – are often cited as a reason for negative energy prices in U.S. electricity markets (\cite{johnston2019nonrefundable}, \cite{aldy2023investment}). The PTC is also credited with helping to increase investment in wind capacity (\citet{wiser2007using}, \citet{osti_950222}, \citet{metcalf2010investment}, \cite{Shrimali2015}); meanwhile, wind generation is associated with negative prices \citep{rafizadeh2025economics}. However, a causal link between the PTC and negative prices has not been established in the literature since wind generation can affect electricity markets through multiple mechanisms. 

This paper contributes to the literature on integrating intermittent renewable generation into wholesale electricity markets and renewable energy subsidies.  We derive an analytical model to encompass multiple mechanisms that account for negative electricity prices and then employ quasi-experimental variation in PTC eligibility to causally estimate the effect of the subsidy on negative electricity prices.  We use these causal estimates to calculate the relative importance of the PTC compared to other drivers of negative prices such as ramping constraints. Our empirical analysis focuses solely on wind generation since large-scale solar penetration into wholesale electricity markets is a relatively recent development and thus not adequately represented in the available data.

To start, we collect hourly price and load data for all five wholesale electricity markets (ISO-NE, MISO, NYISO, PJM, and SPP) operating in the Eastern Interconnection.\footnote{The market acronyms stand for following: ISO-NE = Independent System Operator New England; MISO = Midcontinent Independent System Operator; NYISO = New York Independent System Operator; PJM = PJM Interconnection; and, SPP = Southwest Power Pool.} 
These data span the years 2000-2023 but not all years are available for all markets.\footnote{SPP expanded service into the U.S. Western Interconnection in April 2026.} 
We focus only on the energy component of the electricity price and document that 18 percent of negative energy price hours are below (i.e., more negative) than could be induced by the PTC alone. This provides \textit{prima facie} evidence that a significant share of hours with a negative energy price have a causal driver on the margin other than the federal production tax credit for wind generation. Other mechanisms that can lead to negative energy prices in electricity markets include generator ramping constraints, inter-regional transmission constraints, and state-level subsidies \citep{lopez2021economics}.

Next, we estimate the causal effect of the PTC on the probability of negatively priced energy hours. The exogenous variation to provide the causal estimate is both spatial and temporal. Spatial variation comes from the 10-year limit for PTC eligibility leading to exogenous variation in PTC-eligible wind capacity across markets. Temporal variation comes from the inherent intermittency of wind generation. Given the spatial and temporal variation, our identifying assumption is that the wind resource cannot select across PTC-eligible versus non-PTC eligible wind generators in a given hour. We find that the PTC induces negative energy prices, but the marginal effect is on average small, although statistically significant. Our preferred specification finds that increasing the amount of PTC-eligible wind generation by 1 percent in a given hour increases the probability of a negative energy price by 3.4 percentage points. Furthermore, using dominance analysis, we conclude that only 10 percent of the total variation in negative price hours is explained by the PTC. However, significant heterogeneity in the causal estimate exists across markets; specifically, the effect size in SPP -- an electricity market in the central US plains -- is an order of magnitude larger than average effect across the U.S. Eastern Interconnection. Thus, the PTC has a small effect on negative prices on average, but in specific markets (i.e., SPP) it can have a large effect. 

Finally, we employ our causal estimate to calculate the value of environmental benefits from additional wind generation as a result of the PTC in hours with negative prices. Using marginal environmental damage estimates from \cite{Holland2016}, we calculate that the value from reduced local and global emissions (specifically, CO2, SO2, NOx, PM2.5, VOCs) is \$655-1,092 million in the negatively-priced hours from 2014-2022. These benefits are double the PTC-induced expenditure of \$286-521 million, suggesting the PTC yielded net benefits in these hours. Furthermore, motivated by the missing-market model of \cite{wright1989theory}, our study demonstrates that additional electricity transmission and storage capacity would produce additional total surplus in electricity markets, all else equal. However, since our study can only causally identify the effect during negatively priced energy hours, we cannot calculate a total value from expanding transmission capacity and energy storage across all hours.\footnote{In fact, negative prices in some hours may be optimal if transmission and storage are sufficiently costly.} Regardless, causally showing the existence of these missing markets also informs the the debate on the value of energy storage (\citet{mcconnell2015estimating}, \citet{zhou2016electricity}) and underinvestment in transmission capacity (\citet{Davis2023}, \citet{Hausman2025}).

The next section provides additional background on negative electricity prices and the PTC. Section \ref{Sec:Model} builds an analytical model to motivate the empirical framework in Section \ref{Sec:Empirical}. Meanwhile, Section~\ref{Sec:Data} details the data sources and provides summary statistics while Section \ref{Sec:Results} reports the results. Section \ref{Sec:Welfare} calculates the net-benefit of the PTC as well as discusses the value of missing storage and transmission in the context of negative prices. Section \ref{Sec:Conclusion} briefly concludes.

%%%%%%%%%%%%%%%%%%%%%%%%%%%%%%%%%%%%%%%%%%%%%%%%%%
%%%%%%%%%%%%%%%%%%%%%%%%%%%%%%%%%%%%%%%%%%%%%%%%%%
%%%%%%%%%%%%%%%%%%%%%%%%%%%%%%%%%%%%%%%%%%%%%%%%%%

\section{Background} \label{Sec:Background}

%%%%%%%%%%%%%%%%%%%%%%%%%%%%%%%%%%%%%%%%%%%%%%%%%%
%%%%%%%%%%%%%%%%%%%%%%%%%%%%%%%%%%%%%%%%%%%%%%%%%%

\subsection{Negative Prices in Electricity Markets}

Electricity markets in the U.S. are organized by independent system operators (ISOs) and regional transmission organizations (RTOs). ISOs and RTOs operate regional wholesale electricity markets for a particular geographic region. Operating a wholesale market means an ISO/RTO must ensure that electricity supply perfectly balances with demand.  Without loss of generality, we will refer to these markets as ISOs.

In these electricity markets, the price of electricity is typically known as the locational marginal price (LMP). The LMP measures the wholesale electricity price at specific locations within a regional power grid and it is comprised of three components: (1) energy; (2) congestion (i.e., intra-regional transmission constraints); and (3) losses (i.e., waste heat from transmission lines). The market-wide wholesale energy price is generally determined by an auction mechanism where electricity producers submit price-quantity bids to the ISO resulting in a supply curve. The auction (or market) price clears the market to meet demand. All producers that bid lower than, or equal to, the market clearing price are dispatched to generate electricity and earn the market energy price regardless of their bid. The energy price is a common price across all locations within an ISO. 

As a result, the term ``negative price'' is generally ambiguous in the context of wholesale electricity pricing as a ``negative price'' can mean that either the LMP is negative or one of its components is negative (such as the wholesale energy price). However, a negative LMP does not guarantee a negative energy price, or vice versa. This study focuses on negative energy prices since energy prices reflect the market clearing conditions absent other constraints, revealing the opportunity cost of production.\footnote{In an Arrow-Debreu economy, the assumptions of monotonic preferences and no arbitrage, along with free disposal, guarantee positive equilibrium prices \citep{MasColellWhinstonGreen1995}. However, if the free disposal assumption is relaxed and costly disposal is necessary, then negative equilibrium prices can arise \citep{hart1975proof,repullo1987existence}. The cost to producers of excess electricity supply means free disposal no longer holds. In the wholesale electricity market setting, load-serving entities such as utility companies and retail suppliers are typically the buyers of energy, and thus these are the entities that are ``paid'' when energy price goes negative. The accounting practices with respect to negative prices are not addressed here.} 

The literature identifies two main causal mechanisms that can lead to negative energy prices in an electricity market. The first mechanism is direct production subsidies. As the model in Section~\ref{Sec:Model} demonstrates, a direct production subsidy means that a generator has the incentive to bid below its marginal cost to gain a more favorable position in the dispatch order and increase the probability of earning the subsidy payment. Thus, if a generator has a sufficiently low marginal cost, then it is optimal to bid a negative price. Then, a market may clear at a negative price if there is sufficiently low demand.\footnote{A market-clearing negative price is conceptually different than issue of curtailment (i.e., reducing output from generators). The practice of curtailing occurs to bring a market into equilibrium when supply would exceed demand, but, regardless, a negative energy price can be the market-clearing price.}
In this study, we abstract from market power on both the generator and buyer sides. However, if a generator or set of generators had sufficient market power, then the market price would likely increase, and prior literature finds evidence of implicit collusion in wholesale energy markets (e.g., \cite{mansur2007oligopolists}, \cite{puller2007pricing}).

The second major mechanism leading to negative prices is intertemporal constraints that we refer to as ``ramping'' constraints. This type of constraint arises from thermal generators with boilers powered by nuclear, coal, or natural gas. For these units, future output is conditional on past output as the generator cannot immediately transition (i.e., ``ramp'') from zero generation to full output due to physical limitations. That is, there is a ramping period where these units slowly increase production to reach full capacity, and this can last several hours \citep{denholm2015overgeneration}. Therefore, an operator of a unit with ramping constraints either implicitly or explicitly engages in an intertemporal optimization problem to determine the opportunity cost of not generating in the current period since future revenue is conditional on the current period's output \citep{jha2025start}. This incentive can cause such generators to submit negative bids in the current period to secure future revenue. We provide a model of the inter-temporal mechanism in Section~\ref{Sec:Model} too.

In settings with storage capacity, \cite{wright1989theory} provides a theory of negative prices based on the observation that the price of the commodity for future delivery may fall below the price of immediate delivery. This phenomenon is known as an inverse carrying charge and \cite{wright1989theory} re-characterizes the carrying charge as a negative price of storage.\footnote{In their setting, the two necessary conditions for a negative storage price are as follows: (1) the general commodity is composed of economically distinct sub-aggregates, and (2) the cost to transform one sub-aggregate into another varies across time. For example, assume corn is comprised of corn grown in Iowa and corn grown in Illinois and it is costly to move corn between Iowa to Illinois. This can lead to negative prices in markets with storage due to the differential local, storage, and transportation costs.} Until recently, the short-run storage of electricity has not been feasible or widely available.\footnote{An exception to the historical lack of storage are pump-storage hydroelectric facilities that allow for seasonal storage of electricity by filling up reservoirs with water pumps when electricity is cheap (i.e., Spring and Fall) and then release water to generate electricity when electricity is expensive (i.e. Summer). Net generation from pump-storage hydroelectric is negative over time since the efficiency of the pumps is less than 100\%.  While pumped-storage plants have existed since 1960, they require specific geography and therefore are not particularly common.}
\cite{zhou2016electricity} explicitly models negative electricity prices in a setting with storage (but large-scale grid storage was not available during our study period).

%%%%%%%%%%%%%%%%%%%%%%%%%%%%%%%%%%%%%%%%%%%%%%%%%%
%%%%%%%%%%%%%%%%%%%%%%%%%%%%%%%%%%%%%%%%%%%%%%%%%%

\subsection{Federal Production Tax Credit}\label{Sec:PTC}

The federal renewable electricity Production Tax Credit (PTC) has been the primary production subsidy for renewable electricity generators \citep{sherlock2020renewable}. A stated goal of PTC implementation was the increase in renewable energy capacity and it has been effective at achieving that goal (\citet{wiser2007using}, \citet{osti_950222}, \citet{metcalf2010investment}, \cite{Shrimali2015}).  The subsidy is a per megawatt-hour (MWh) tax credit. The tax credit is applied to all output from a qualifying unit for the 10 years following the beginning of operation. Once a generator satisfies the eligibility requirements for the PTC, then the 10-year credit is effectively irrevocable.\footnote{Initially enacted in 1992, the PTC has been a series of temporary policies passed by Congress. This has led to periods of the PTC lapsing for new units, but the PTC has always been retroactively applied upon future authorization. The One Big Beautiful Bill (OBBB) of 2025 did not immediately end the PTC but initiated a phase out of new PTC eligibility.}

\begin{figure}
    \centering
    \includegraphics[width=1\linewidth]{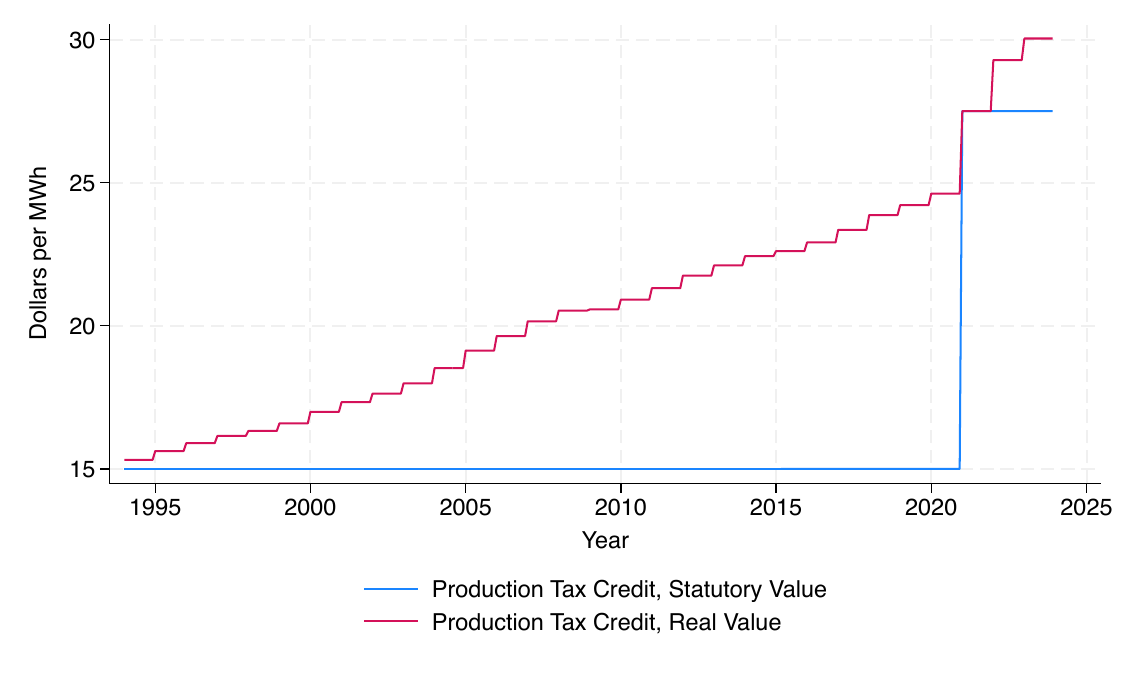}
    \caption{Production Tax Credit Value}
    \label{fig:PTCrate}
\end{figure}

Figure \ref{fig:PTCrate} graphs the statutory value of the PTC by year. The statutory value of the PTC remained constant at \$15/MWh from 1992 to 2021. However, the PTC was indexed to inflation, so the red line provides the real value of the subsidy. The Inflation Reduction Act of 2022 significantly increased the statutory PTC value (and the new level was also index to inflation). By 2024, the real value of the PTC reached \$30/MWh. 

%%%%%%%%%%%%%%%%%%%%%%%%%%%%%%%%%%%%%%%%%%%%%%%%%%
%%%%%%%%%%%%%%%%%%%%%%%%%%%%%%%%%%%%%%%%%%%%%%%%%%
%%%%%%%%%%%%%%%%%%%%%%%%%%%%%%%%%%%%%%%%%%%%%%%%%%

\section{Model} \label{Sec:Model}

This section has three purposes. First, we demonstrate how per unit subsidies create the incentive for generators to bid below their marginal cost leading to negatively-priced bids in a wholesale auction. We focus on the federal production tax credit (PTC) but other federal, state, and/or local subsidies may be available to a producer in addition to the PTC. Second, we clarify how negative bids from generators form the wholesale electricity supply curve and can lead to observed prices that are below zero when demand is sufficiently low. Third, we show that other mechanisms can plausibly produce negative bids. We provide a model of generator ramping in which a generator’s production in future periods is dependent on production in previous periods. Thus, if the expected future price is sufficiently high, then a generator would be willing to bid a negative price in the current period to ensure access to future revenue. Future demand will serve as an important control for our regression analysis.

%%%%%%%%%%%%%%%%%%%%%%%%%%%%%%%%%%%%%%%%%%%%%%%%%%
%%%%%%%%%%%%%%%%%%%%%%%%%%%%%%%%%%%%%%%%%%%%%%%%%%

\subsection{General Model Setup}

Let $q_{it}$ be the quantity supplied by generator $(i)$ in period $(t)$. Next, assume a non-negative constant marginal cost across time for each generator denoted by $c_i\geq0$. Finally, each generator also has a maximum output or capacity constraint given by $\bar q_i\geq0$. The \textit{ex-ante} profit function is then given by $\pi_i=\sum_t \left[(b_{it}-c_i)q_{it}\right]$, where $b_{it}$ is the price bid by the generator into the wholesale market.  The profit function is \textit{ex-ante} because the \textit{ex-post} profit only arises once the market price $p_t$ is set by the ISO following an auction.\footnote{In many cases, output from a wind generator is covered by a power purchasing agreement (PPA) that guarantees a predetermined positive price for the generator's electricity \cite{mesa2023long}. Denote the PPA price as $\bar p_{it}$ since it is set at the generator level but may change over time. If $\bar p_{it}>c_i$, then a wind generator always earns a positive producer surplus when generating and thus we can focus on bids directly without needing to explicitly model a shutdown condition.}
To abstract from market power issues, we assume that each generator is owned by a different profit-maximizing firm.

The model does not have storage, so a non-negativity constraint is included as $q_{it}\geq0$. Utility-scale storage increases demand and thus the price when the energy prices would otherwise be low, so storage can only make negative energy prices less likely.

\begin{figure}
     \centering
     \begin{subfigure}[b]{0.67\textwidth}
         \centering
         \includegraphics[width=\textwidth]{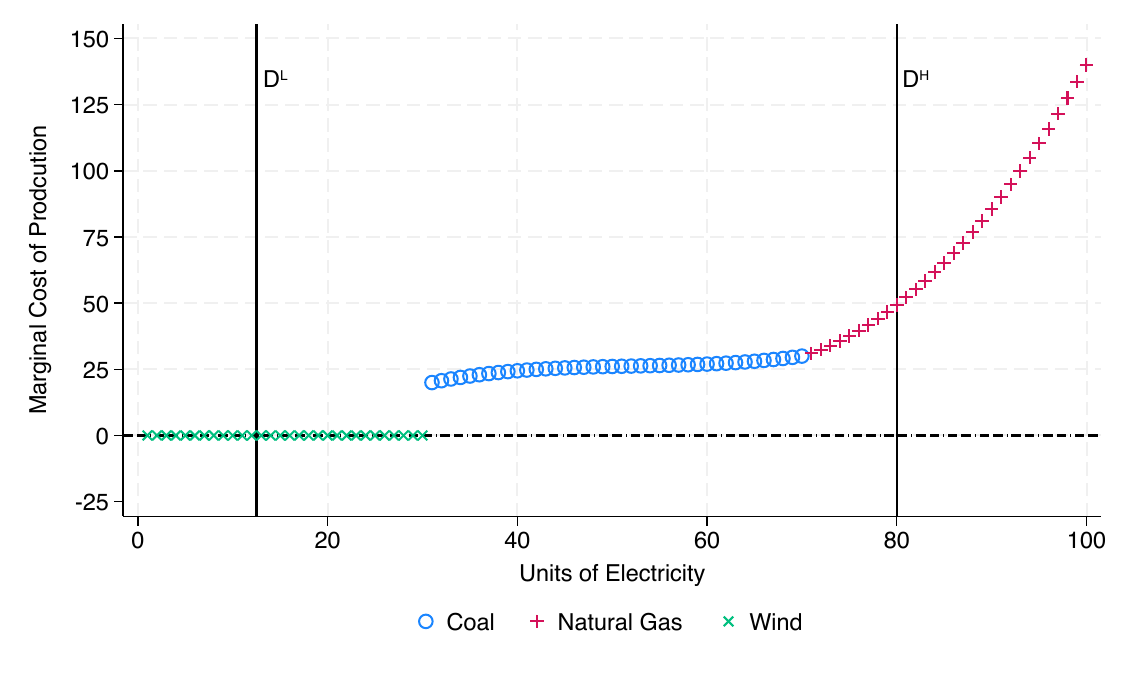}
         \caption{Simple Supply Curve}
         \label{fig:stack}
     \end{subfigure}
     \vspace{1em}
     \begin{subfigure}[b]{0.67\textwidth}
         \centering
         \includegraphics[width=\textwidth]{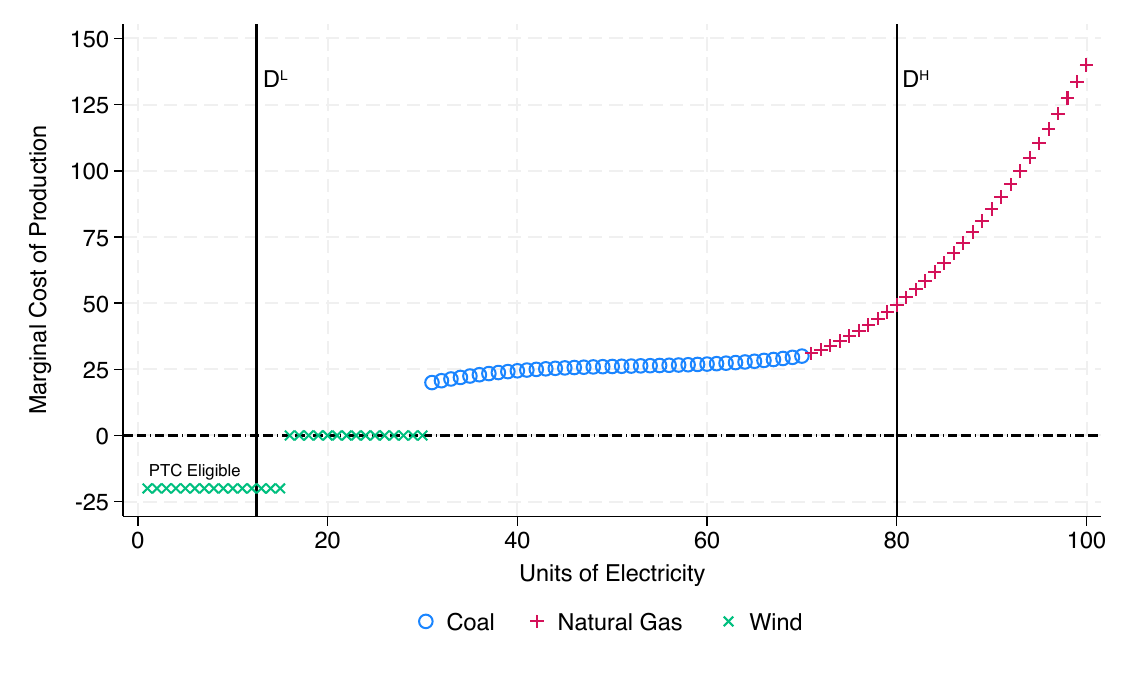}
         \caption{PTC Subsides}
         \label{fig:stack_ptc}
     \end{subfigure}
     \vspace{1em}
     \begin{subfigure}[b]{0.67\textwidth}
         \centering
         \includegraphics[width=\textwidth]{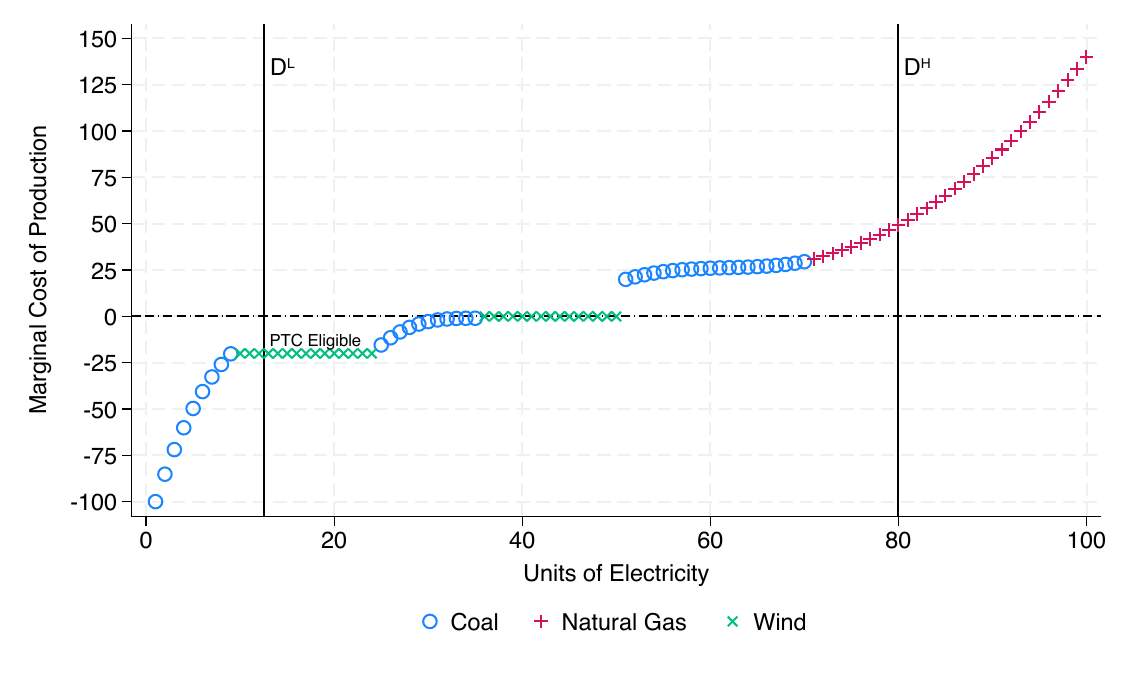}
         \caption{PTC Subsidies and Dynamic Constraints}
         \label{fig:stack_ramp}
     \end{subfigure}
        \caption{Stylized Electricity Sector Supply Curves}
        \label{fig:allstack}
\end{figure}

To fix ideas, Figure \ref{fig:allstack} displays stylized electricity sector supply curves where generators face the optimization problems in our analytical model.  Then, Figure \ref{fig:stack} depicts a stylized single-period wholesale electricity market supply curve with three different types of generating units. The figure assumes that all wind generators have a zero marginal cost of production. The coal generators have differentiated marginal costs but those costs are always above zero but less than the marginal cost of the natural gas generators. The natural gas generators have the highest marginal cost. Ordering the units from lowest to highest marginal cost yields a supply curve. For Figure \ref{fig:stack}, in a low-demand hour $(D^L)$ the market-clearing price is zero, but in a high-demand hour $(D^H)$ the market would clear at a positive price.\footnote{The stylized demand curves are perfectly inelastic to represent the small short-run elasticity of electricity demand.}

%%%%%%%%%%%%%%%%%%%%%%%%%%%%%%%%%%%%%%%%%%%%%%%%%%
%%%%%%%%%%%%%%%%%%%%%%%%%%%%%%%%%%%%%%%%%%%%%%%%%%

\subsection{Subsidies} \label{Subsec:Subsidies}

We turn our attention to the effect of per unit subsidies on generator bids. Let $s_{it}>0$ be the total subsidy that generator $(i)$ in period $(t)$ receives per unit of output. If the PTC is the only subsidy available to a wind generator, then $s_{it}=PTC>0$. The one-period profit maximization problem leads to the following Lagrangian:
\begin{align*}
    \max_{\{q_{it},\lambda,\delta\}} \mathcal{L}=(b_{it}+s_{it}-c_i)q_{it}+\lambda(\bar q_i - q_{it})+\delta q_{it}
\end{align*}
where $q_{it}$ is the choice variable while $\lambda,\delta\geq0$ are Lagrange multipliers. The multiplier $\lambda$ is the shadow price of capacity constraint and $\delta$ relates to the non-negativity constraint. Maximizing over the quantity implicitly determines the optimal bid, where a price-quantity pair $(b_{it},q_{it})$ is submitted to the wholesale market operator.

The Kuhn-Tucker conditions yield $b_{it}=(c_i+\lambda-s_{it}-\delta)$, $\lambda(\bar q_i-q_{it})=0$, and $\delta q_{it}=0$, where $\lambda,\delta \geq 0$. If the capacity constraint binds such that the generator bids its total capacity, then $\lambda>0$ and $q_{it}=\bar q_i$ (implying $\delta=0$ as $q_{it}>0$). In this case, $b_{it}=c_i+\lambda-s_{it}$, where we interpret $\lambda>0$ as the additional cost of running a unit beyond its nameplate capacity.\footnote{Typically, the nameplate capacity of a wind generator is the output in megawatts at ideal conditions, but conditions may deviate from the ideal due to the age of the turbine and excess wind speed leading to the possibility of an ``overrun'' condition.} 
If the capacity constraint just binds for wind generators, then $\lambda=0$ with $q_{it}=\bar q_i$. The multiplier $\lambda$ being slack means that the inequality $q_{it} \leq \bar q_i$ satisfies the first-order conditions. However, it is profit maximizing for a generator bid all its capacity in this one-period optimization if the generator submits a positive quantity bid, and thus $q_{it}=\bar q_i$, always. Observe, if $\delta>0$, then $q_{it}=0$, but any positive gross price and positive quantity bid yields positive short-run profit and that dominates a zero-quantity bid that guarantees zero profit, and therefore $\delta=0$.

Therefore, a generator submits a negative bid when the following holds: $s_{it}>c_i+\lambda$. That is, the bid is negative if the subsidy is large enough to overcome both the marginal cost and the capacity constraint multiplier. A subsidy less than the marginal cost is not sufficient to generate a negative bid. However, if a wind generator has zero marginal cost per unit of output $(c_i = 0)$ and the capacity constraint is slack $(\lambda =0)$, then any subsidy leads to a negative bid. The model also provides the lowest possible bid given a subsidy structure. If $c_i=\lambda=0$, then the bid is given by $b_{it}=-s_{it}<0$. Therefore, if the production tax credit is the only subsidy to a generator, then the minimum possible bid is $b_{it}=-PTC$. Generally, we can rewrite the total subsidy as $s_{it}=(PTC+OTHER)$ where $OTHER\geq0$ is the sum of all other federal, state, and/or local production subsidies.\footnote{The most prevalent other production subsidy is from renewable portfolio standards where generators produce and sell renewable energy credits (RECs) for compliance with the portfolio standard.} Then, the minimum possible bid can be rewritten as $b_{it}=-(PTC+OTHER)<0$, and this expression emphasizes that any bid lower than the negative of the PTC must come for another subsidy.

Figure \ref{fig:stack_ptc} builds on the foundations of Figure \ref{fig:stack}, but adds the effect of PTC subsides. Now, the PTC-eligible wind generators have a negative marginal cost due to the opportunity cost of the subsidy. The wind generators not eligible for the PTC remain at a zero marginal cost. All of the coal and natural gas generator costs are unaffected by the renewable energy subsidy. Here, the demand curve $D^L$ intersects the supply curve below zero leading to a negative energy price. This highlights that negative energy prices are a function of both demand and supply. For Figure \ref{fig:stack_ptc}, the negative price is exactly equal to $-PTC$.

%%%%%%%%%%%%%%%%%%%%%%%%%%%%%%%%%%%%%%%%%%%%%%%%%%
%%%%%%%%%%%%%%%%%%%%%%%%%%%%%%%%%%%%%%%%%%%%%%%%%%

\subsection{Ramping} \label{Subsec:Ramping}

Here, we show that negative energy prices can arise via another mechanism in addition subsidies. Ramping is the phenomenon where a generator cannot instantaneously switch from zero output to generating at full capacity. This occurs in large thermal electric generators and can take several hours to reach full capacity \citep{kirby2005method}. In this version of the model, generators subject to a ramping constraint may bid negative prices in the current period to ensure operation at a higher output level in a later period.

For illustrative purposes, we build a two-period model with $t=1,2$ where the generator can optimize bids over both periods. The capacity and non-negative constraints still apply (as in the subsidy model above), but a new ramping constraint given by $q_{i2} - q_{i1} \leq \bar d_i$ is included where $\bar d_i>0$ is a generator-specific maximum change in output from one period to the next. The cost structure is unchanged, and so the two-period profit maximization problem with a ramping constraint yields to following Lagrangian:
\begin{align*}
    \max_{\{q_{i1},q_{i2},\lambda_1,\lambda_2,\delta_1,\delta_2,\gamma\}} \mathcal{L} =& (b_{i1}-c_i)q_{i1}+(b_{i2}-c_i)q_{i2}+\lambda_1(\bar q_i - q_{i1})+\lambda_2(\bar q_i - q_{i2}) \\
    &+\delta_1 q_{i1} +\delta_2 q_{i2} + \gamma (\bar d_i +q_{i1}-q_{i2})
\end{align*}
where $\gamma \geq 0$ is the shadow price on the ramping constraint. It must be that the ramping constraint binds for the ramping mechanism to generate a negative bid, so we limit our analysis to cases where $\gamma>0$. Logically, if the bid in the first periods is less than or equal to zero $(b_{i1} \leq 0)$, then the bid in the second period must be greater than or equal to zero $(b_{i2} \geq 0)$ since the generator can choose not to bid in both periods and earn zero profits as the outside option. When the ramping constraint binds, then $q_{i1}<\bar q_i$ and $\lambda_1=0$, and thus we analyze the non-trivial case where $\lambda_1=0$. In this case, the optimal bid is $b_{i1}=c_i-\gamma$ for some $0<q_{i1}$, and clearly the bid is negative when $\gamma>c_i\geq0$. This shows that negative bids can arise from another mechanism in the electricity sector besides direct subsidies such as the federal production tax credit. 

The ramping constraint multiplier, $\gamma$, can be large and lead to negative prices below the minimum bid that PTC can induce (as described above). Such negative prices are plausibly found in the data as demonstrated in Section \ref{Sec:Data}. For our regression specifications and results in Sections \ref{Sec:Empirical} and \ref{Sec:Results}, respectively, we employ a vector of future load values to capture the multi-hour ramping process at many thermal electricity generators.

Figure \ref{fig:stack_ramp} then builds on Figure \ref{fig:stack_ptc} by adding the effects of dynamic optimization to the single-period supply curve. For illustrative purposes, we assume that dynamic optimization affects only coal-fired generators due to their long ramping times relative to natural gas generators. Then, we depict a range of ramping effects where some units are unaffected and still have a positive marginal costs. However, some units have very large ramping constraints (i.e., $\gamma>>0$) leading to a real marginal cost that is more negative than the PTC-eligible wind generators (see far left of the supply curve). Other dynamically constrained coal-fired generators experience a small negative marginal cost and thus fit between the two segments of wind generators. In this case, $D^L$ still intersects with the supply curve such that the market price is $-PTC$, but other negative price values are possible if the demand was slightly lower or higher.

%%%%%%%%%%%%%%%%%%%%%%%%%%%%%%%%%%%%%%%%%%%%%%%%%%
%%%%%%%%%%%%%%%%%%%%%%%%%%%%%%%%%%%%%%%%%%%%%%%%%%

\subsection{Equilibrium Prices} \label{Subsec:Profits}

How do negative bids turn into negative energy prices? Let $Q_t^D$ be the exogenous quantity demanded in period $t$ for a given wholesale electricity market (recalling that the energy portion of a LMP is determined at the aggregate market level).\footnote{Alternatively, we can allow the quantity demanded to be a function of price denoted as $Q_t^D(p_t)$ with $dQ_t/dp_t<0$, but the supply curve is flat in the PTC-eligible region and thus the slope of demand is irrelevant with respect to the existence of negative prices.} The goal of the market operator is to set a market price that balances demand and supply at the lowest possible cost, and does so by creating a supply curve from the generators bids and then finding the market-clearing price given the exogenous demand.

For simplicity of exposition, we assume that a wholesale electricity market is uniform (i.e., one pricing zone) and there are N generators. In each period, the market operator receives the complete set of bids $B_t={(b_{1t},q_{1t} ),\ldots,(b_{Nt},q_{Nt} )}$. Then, the operator sets a market clearing price $p_t$ such that the quantity demanded equals quantities supplied meaning $Q_t^D=Q_t^S \equiv \sum_i [q_{it}\cdot \mathds{1}(b_{it}\leq p_t)]$, where $\mathds{1}(b_{it}\leq p_t)$ is an indicator function taking the value 1 if the generator's bid is less than or equal to the market clearing price, and zero otherwise. We abstract from the lumpiness of the $q_{it}$ values such that $Q_t^D=Q_t^S$ without additional notation.\footnote{In practice, balancing authorities enable quantity demanded to equal quantity supplied through so-call ``spinning reserves'' for thermal generators and curtailments for renewable generation such as wind and solar.} All bids that satisfy the condition $b_{it}\leq p_t$ then generate electricity in the period. That is, generators with bids higher than the market clearing price have an \textit{ex-post} quantity supplied of zero and earn no revenue. 

The market clearing energy price is negative if there is sufficient capacity supplied from generators with negative bids to meet demand. Thus, a negative price in a particular period is more likely given one or more conditions. The first condition is the overall share of capacity in an ISO with subsidies is large. In the limit, if all the capacity in an ISO was from PTC-eligible wind generators, then all the bids would be negative and there would always be a negative energy price. At the other extreme and ignoring ramping constraints, if there were no generator-level subsidies within an ISO, then all bids would be positive, and negative prices would never be observed. Therefore, it follows that the prevalence of negative energy prices in positively related to the share of capacity from PTC-eligible wind generators. The second condition is that quantity demanded is low. This means the demand curve is more likely to hit the negative part of the supply curve when such a negative segment exists. The third condition is as follows: if the prevalence of generators with ramping constraints increases, then it is more likely some generators in an ISO will find it profitable to bid a negative price in a given hour, all else equal.

%%%%%%%%%%%%%%%%%%%%%%%%%%%%%%%%%%%%%%%%%%%%%%%%%%
%%%%%%%%%%%%%%%%%%%%%%%%%%%%%%%%%%%%%%%%%%%%%%%%%%
%%%%%%%%%%%%%%%%%%%%%%%%%%%%%%%%%%%%%%%%%%%%%%%%%%

\section{Data} \label{Sec:Data}

We employ data on electricity prices from all the competitive wholesale electricity markets in the U.S. Eastern Interconnection to estimate the causal impact of the federal PTC for wind generation on the frequency of negative energy prices. We combine this market data with information on each wind generator to determine its PTC eligibility. We also provide \textit{prima facie} evidence that a significant share of hours with negative energy prices have a causal driver on the margin other than the federal PTC.

%%%%%%%%%%%%%%%%%%%%%%%%%%%%%%%%%%%%%%%%%%%%%%%%%%
%%%%%%%%%%%%%%%%%%%%%%%%%%%%%%%%%%%%%%%%%%%%%%%%%%

\subsection{Sources}

%%%%%%%%%%%%%%%%%%%%%%%%%%%%%%%%%%%%%%%%%%%%%%%%%%

\subsubsection{Electricity Prices and Demand}

We collect data on hourly electricity prices and electricity demand (load) for all five of the competitive wholesale electricity markets in the U.S. Eastern Interconnection: ISO-New England (ISO-NE), Midcontinent Independent System Operator (MISO), New York Independent System Operator (NYISO), the Pennsylvania, New Jersey, and Maryland regional transmission organization (PJM), and Southwest Power Pool (SPP).\footnote{The Southeast portion of the United States does not have a competitive wholesale electricity market with transparent prices. Instead, electricity is sold through bilateral trades and therefore we do not have information on either hourly prices or demand.} 
Each of these ISOs publishes detailed hourly data on total load along with prices and price components (energy, congestion, and losses) at each pricing point (i.e., zone) in their territory. Data from ISO-NE spans the years 2007-2022, MISO spans 2005-2022, NYISO spans 2000-2023, PJM spans 2007-2022 and SPP spans 2014-2023. Overall, this gives us 728,128 hours of data across all ISOs in our sample. As discussed in Section \ref{Sec:Background}, we focus on the energy price component that is uniform across an ISO at any given time. 

\begin{figure}
    \centering
    \includegraphics[width=1\linewidth]{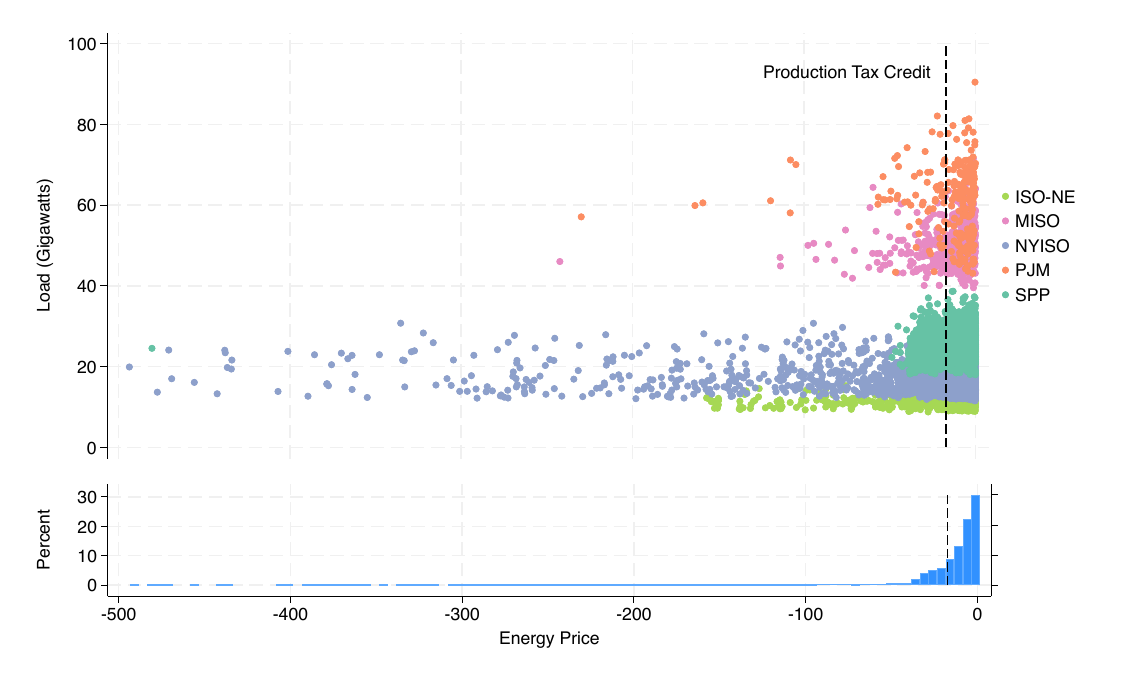}
    \vspace{6pt} % Optional small gap
    \begin{minipage}{0.8\textwidth}
    \footnotesize Note: Figure excludes 43 prices smaller than -\$500.
    \end{minipage}
    \caption{Price Distribution of Negative Energy Prices and Load by ISO}
    \label{fig:PriceDistribution}
\end{figure}

To visualize the distribution of negative energy prices, the top panel of Figure \ref{fig:PriceDistribution} provides a scatter plot of all negative electricity prices and load, color coded by ISO.  The dashed line near the right hand side of the figure displays the average value of $-PTC$ over our sample period.\footnote{We display the average value here since, as discussed in Section \ref{Sec:Background}, the real value of the PTC varies over time.}  While many negative prices are between zero and the (negative) PTC, a substantial portion of negative prices are more negative than the PTC. Moreover, it is clear that negative prices are somewhat common phenomenon across all markets. The bottom panel of Figure~\ref{fig:PriceDistribution} shows a frequency distribution of negative prices. The data indicate that nearly 18\% of hours with negative energy prices have prices more negative than the value of the PTC, implying that the PTC cannot be the only reason for observing negative prices on the margin. This supports the modeling in Section \ref{Sec:Model} showing multiple economic channels for negative prices including intertemporal ramping constraints.

\begin{figure}
    \centering
    \includegraphics[width=1\linewidth]{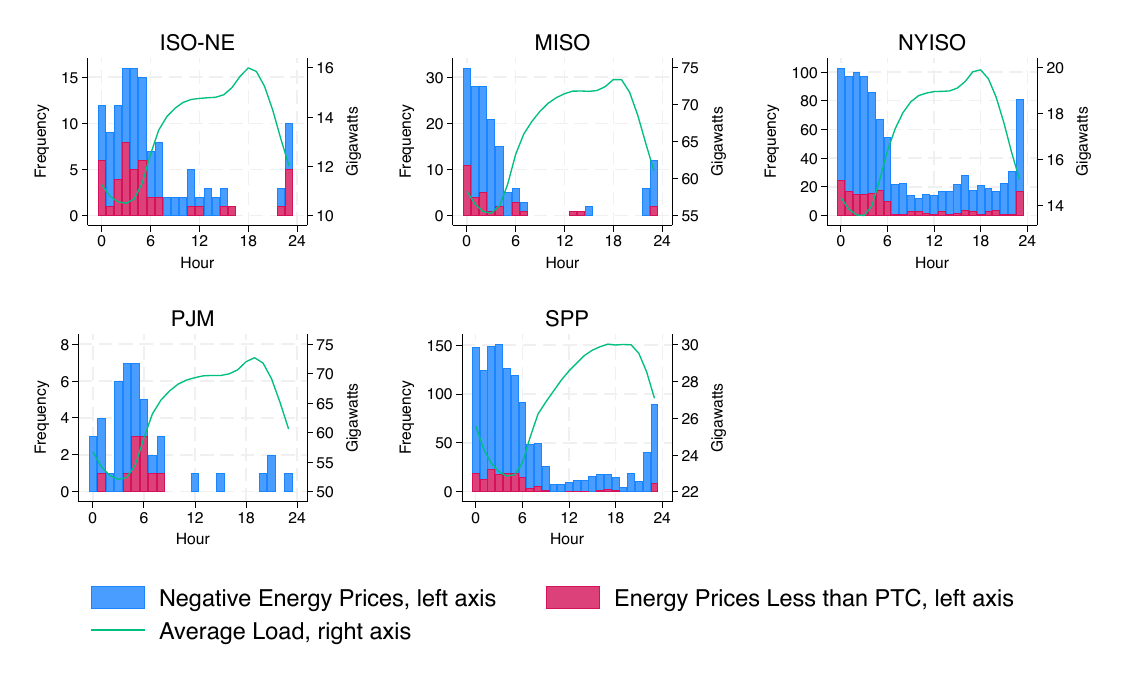}
    \caption{Negative Energy Price Hours and Average Hourly Load in Autumn}
    \label{fig:PricesAutumn}
\end{figure}

We also observe in our data how ramping constraints might be causing negative prices. Figure~\ref{fig:PricesAutumn} shows the frequency of negative prices by hour of the day with the average hourly load in an ISO overlaid for hours in the autumn months of September, October, and November. To start, the figure demonstrates that hours with low electricity demand are much more likely to have negative electricity prices as our model predicts (see blue bars). Next, Figure~\ref{fig:PricesAutumn} also provides preliminary evidence that negative energy prices less than the PTC (see red bars) are more likely to occur in the hours preceding an increase in demand.\footnote{In Figure~\ref{fig:PricesAutumn}, the red bars are a subset of the blue bars. The data are split by season due to different hourly demand patterns across seasons. However, a similar pattern is observed in other seasons.}
This suggests that intertemporal constraints on generators may play an important role in prevalence of negative prices. The empirical specifications provided in Section \ref{Sec:Empirical} employ a flexible function of future load to help predict the probability of negative energy prices in the current period.

%%%%%%%%%%%%%%%%%%%%%%%%%%%%%%%%%%%%%%%%%%%%%%%%%%

\subsubsection{Wind Generator Eligibility for Production Subsidies}

We merge our market data with information on wind generator eligibility to receive the federal Production Tax Credit (PTC). To start, we use data from Form EIA-860 that contains detailed information on every electricity generator with a capacity of 1 megawatt or larger in the United States. The EIA-860 data include the fuel that each generator uses as well as the month and year each began operating. Combining this information with a legislative history of the PTC, see Section~\ref{Sec:PTC}, allows us to determine the amount of wind capacity eligible for the PTC recalling a generator receives the subsidy for ten years from its in service date. 

\begin{figure}
    \centering
    \includegraphics[width=1\linewidth]{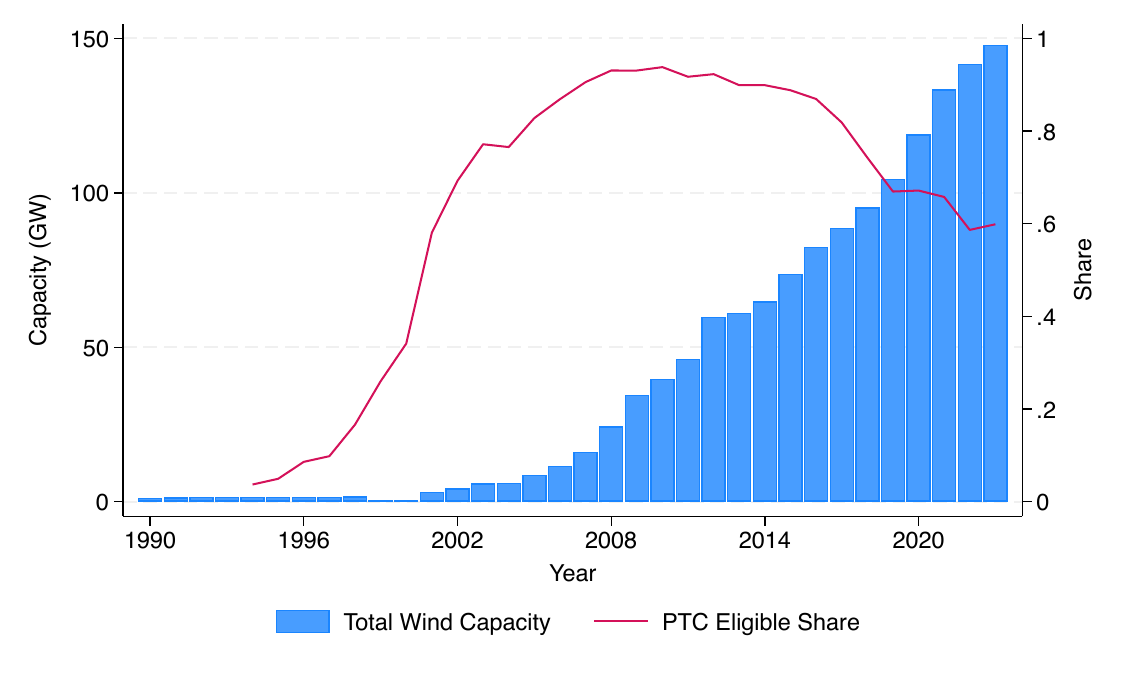}
    \caption{U.S.\ Wind Capacity and PTC Eligible Share by Year}
    \label{fig:WindCapByYear}
\end{figure}

As Figure \ref{fig:WindCapByYear} shows, the absolute amount of wind generation capacity has increased dramatically over the last three decades from 1.4 GW in 1990 to 149 GW in 2023. The red line on the figure (right axis) displays the fraction of the wind generation that is eligible to receive the PTC in that year. The share of wind that was eligible for the PTC grew rapidly from 1997 onward, three years after the PTC was originally passed in 1994, and peaked in 2010 at 94 percent of wind generators. After 2010, the share of PTC eligible wind generation began to decline as generators exhausted their 10-year eligibility window for the PTC, and fell to 60 percent of generators being eligible for the PTC in 2023. We take advantage of this quasi-experimental variation in eligible to help identify the effect of the PTC on negative energy prices.

We combine this PTC eligibility with two additional data sources to impute hourly wind generation from PTC eligible generators.  We take monthly, facility-level, wind generation data from Form EIA-923 and hourly wind generation within an ISO from fuel-mix data reported by each ISO.\footnote{Our approach distinguishes each addition to a wind facility, effectively treating any expansion as a new generator.  For instance, if a wind facility began operation of 10MW in 1997 and then added another 10MW of capacity in 2005, we treat the first 10MW of being PTC eligible until 2007 and the second 10MWs being PTC eligible until 2017.}

To calculate the amount of PTC eligible wind generation in each hour, we first match our data on a generator’s PTC eligibility with its monthly electricity production as reported on EIA-923.  This matching allows us to determine how much wind generation in a particular month is eligible for PTC payments.  

Then, since wind generation is not randomly distributed throughout a month and is subject to variation due to weather conditions, we approximate hourly wind generation using aggregate wind generation data from each ISO.  To allocate a wind plant’s monthly reported generation to a particular hour of the month, we assume that each generator produces electricity whenever the wind is sufficiently strong.  Since wind generators have a marginal cost of approximately zero this is a reasonable assumption. To produce values for hourly generation by each generator, we first compute the share of monthly wind generation for each hour of the month within an ISO to the total monthly wind generation in that ISO.  We then allocate the monthly wind generation from each generator as reported on EIA-923 in proportion to the total wind share in the ISO in each hour.  Equation~\eqref{eqn:windshare} shows this mathematically for wind generator $i$, in hour $h$, in month $m$, and in ISO $r$.
\begin{align}
\textrm{Wind Generation}_{ihmr} &= \frac{\textrm{ISO Generation}_{hmr}}{\sum_{h=1}^H\textrm{ ISO Generation}_{hmr}} \times \textrm{Wind Generation}_{imr}
\label{eqn:windshare}
\end{align}
where $H$ is the total number of hours in a month. This methodology assumes that each wind plant in an ISO is exposed to the same wind conditions in each hour. Although stylized, all errors introduced into our regression estimates via this method can be described as classical measurement error because wind generators cannot strategically choose their production due to the exogeneity of the wind resource.

%%%%%%%%%%%%%%%%%%%%%%%%%%%%%%%%%%%%%%%%%%%%%%%%%%
%%%%%%%%%%%%%%%%%%%%%%%%%%%%%%%%%%%%%%%%%%%%%%%%%%

\subsection{Summary Statistics}

Table \ref{Tab:SummaryStats} shows information about the hourly data we use in our analysis spanning 2001 to 2023. Table~\ref{Tab:SummaryStats} Panel A displays information for all hours across all ISOs. This panel shows that wholesale energy prices vary widely across all hours with an average price of about \$40 per MWh, but with a standard deviation of \$46. In addition, approximately 1.6 percent of hours during that time period exhibit negative prices with 18 percent of those hours exhibiting negative prices are smaller than the Production Tax Credit (0.3 percent of total hours in our sample).

\begin{table}[!htpb]
    \centering
    \caption{Summary Statistics (nominal dollars)} \vspace{1mm}
    \begin{tabular}{rccccc}
    \hline \hline 
       & Mean & SD & Min & Max & N \\
    \hline
    \multicolumn{6}{c}{\textbf{Panel A: All Hours}} \\
    \hline
    Energy price (USD/MWh) & 39.69 & 45.93 & -2838.43 & 4668.08 & 719,344\\
    PTC rate (USD/MWh) & 22.96 & 2.90 & 17.33 & 30.04 & 719,344\\
    Negative energy price indicator & 0.016 & 0.124 & 0 & 1 & 719,344\\
    Energy price in [-PTC,0] indicator & 0.013 & 0.112 & 0 & 1 & 719,344\\
    Energy price is less than PTC indicator & 0.003 & 0.052 & 0 & 1 & 719,344\\
    Hourly load (MW) & 41,499 & 29,884 & 0 & 158,071 & 795,359\\
    Hourly wind generation (MW) & 3,736 & 4,280 & -18 & 24,482 & 490,819\\
    Percent of load met by wind & 8.54 & 11.52 & -0.024 & 87.54 & 446,116\\
    \hline
    \multicolumn{6}{c}{\textbf{Panel B: Negative Energy Price Hours}} \\
    \hline
    Energy price (USD/MWh) & -20.64 & 74.76 & -2838.43 & 0 & 11,153\\
    Energy price is less than PTC indicator & 0.176 & 0.381 & 0 & 1 & 11,153\\
    Hourly load (MW) & 22,328 & 10,944 & 8,784 & 90,489 & 9,917\\
    Hourly wind generation (MW) & 7,592 & 7,087 & 0 & 22,550 & 8,988\\
    Percent of load met by wind & 28.28 & 25.34 & 0 & 87.51 & 7,899\\
    \hline \hline
    \end{tabular}
    \label{Tab:SummaryStats}
\end{table}

As our model predicts, hours with negative prices are observably different than hours with positive prices. Table \ref{Tab:SummaryStats} Panel B contains information only on hours with negative prices.  Comparing Panels A and B shows that hours with negative energy prices have approximately half the load as the full sample on average. As modeled in Section \ref{Sec:Model}, hours with low demand will have lower prices and a higher likelihood of negative prices, all else equal. We also see that hours with negative energy prices have more than twice the wind generation of the average hour in the sample (3,736 MW compared to 7,592 MW). The combination of more wind production along with wind generation being correlated with low-demand periods (i.e., nighttime) leads to a much greater likelihood of negative prices. 

Among the hours with negative energy prices, the average energy price was -\$20.64; meanwhile, the PTC averaged \$22.96 in the sample period. However, the full distribution of negative prices plotted in the bottom panel of Figure \ref{fig:PriceDistribution} demonstrates a skewed distribution with 18 percent of hours more negative than the PTC. Every ISO in our sample has a substantial number of hours when prices are more negative than the PTC, suggesting that greater wind generation cannot be solely responsible for negative prices. 

We compute the total implied payments made by electricity generators to produce electricity over our sample time period.  Multiplying the load by the negative price in each hour yields an estimate of \$4.5 billion (nominal) dollars paid by generators.  Although only 18 percent of hours with negative prices are lower than the PTC, these hours account for 69 percent of payments made by generators (\$3.1 billion). In contrast, the 82 percent of hours where the energy prices are between zero and the PTC value account for 31 percent of the payments: \$1.4 billion.  Our goal is to disentangle the effect of the PTC and other generator constraints with the electricity price between $-PTC$ and zero.

In the next section, we describe our empirical strategy to estimate the causal effect of the PTC on the probability of observing negative prices. However, we must constrain our sample to hours that have data for energy price, electricity demand, and hourly wind generation within an ISO. This restricts our sample to more recent years when the ISOs began publishing the hourly fuel mix used to meet demand within the ISO. Thus, the earliest data in our hourly regression analysis come from 2008, and leaving us with a sample of 419,640 hours with all three necessary data components.

%%%%%%%%%%%%%%%%%%%%%%%%%%%%%%%%%%%%%%%%%%%%%%%%%%
%%%%%%%%%%%%%%%%%%%%%%%%%%%%%%%%%%%%%%%%%%%%%%%%%%
%%%%%%%%%%%%%%%%%%%%%%%%%%%%%%%%%%%%%%%%%%%%%%%%%%

\section{Empirical Framework} \label{Sec:Empirical}

The goal of our identification strategy is to causally measure the effect of the PTC on the prevalence of observed negative energy prices in electricity markets. As we focus on wind generators for identification, then the ideal experiment would be to randomly assign per unit energy subsidies equal to the value of the PTC to wind generators located in random ISOs and then statistically evaluate the change in the number of negative energy price hours. However, randomizing the subsidies did not occur explicitly. Instead, our identification strategy fundamentally relies on variation created by the 10-year limit on PTC eligibility once a wind generator became operational. As discussed above, our dataset contains the month that a wind generator became operational, and thus the 10-year rule creates variation by month in the capacity of PTC-eligible wind generators. That is, spatial variation arises as the share of PTC-eligible wind capacity relative to total wind capacity varies across ISOs within the same time period. This spatial variation in the share of PTC-eligible wind allows us to separate the effect of PTC production subsidies from contemporaneous shocks such as fuel price changes for thermal generators.

As Section \ref{Sec:Model} argues, the prevalence of negative energy prices is positively related to the share of PTC-eligible wind capacity within an ISO. Formally, our identifying assumptions for causal inference are that (i) the wind resource cannot select between PTC-eligible and non-PTC eligible generators within a region and (ii) the 10-year rule means that PTC eligibility determined a decade earlier is not related to unobservables that impact the current energy price. This identification strategy is reinforced by hourly variation in wind generation -- independent of PTC eligible capacity -- due to the natural variability of wind resources; that is, on an hourly basis the amount of wind is exogenous.

\vspace{3mm}
\noindent\textit{Main Specification.} Let $NegPrice=1$ when an ISO's hourly energy price is greater than the negative PTC value but below zero, and 0 otherwise. This binary dependent variable focuses on the interval $[-PTC,0)$ to isolate the impact of PTC on negative prices. As motivated by the model in Section \ref{Sec:Model}, let the main explanatory variable be denoted $WindGenPTC$ and equal to generation in megawatt-hours from PTC-eligible wind generators in ISO $(i)$, hour $(h)$, month $(m)$, and year $(y)$. Then, for this binary dependent variable, we proceed with a logistic regression given by:\footnote{For our preferred specification reported in Section~\ref{Sec:Results} , the mean predicted probability from the logit model is 0.0174 with a standard deviation of 0.0533, and the density is heavily concentrated near zero. Thus, a linear probability model is unsuitable for our setting and produces unreliable estimates. Therefore, we rely on logit estimates as our preferred specification. The average marginal effects are similar if instead we use a probit model and thus indicating that our results are not sensitive to the choice between logistic and normal error distributions.}
\begin{equation} \begin{split}
    Pr(NegPrice_{ihmy}=1|\textbf{X}) =& \Lambda ( \alpha+\beta \times WindGenPTC_{ihmy}+\gamma \times WindGenNonPTC_{ihmy} \\
    &+ f(Load_{i0},\ldots,Load_{i6}) + \omega_{my} + \omega_{hm} + \delta_{iy} + \delta_{im} + \delta_{ih}+ \epsilon_{ihmy} ) \label{Eq:MainSpecification}
\end{split} \end{equation}
where $\Lambda$ is the logistic function (and $\textbf{X}$ denotes the explanatory variables in equation~\eqref{Eq:MainSpecification}). In addition to the main explanatory variable, other control variables include wind generation from non-PTC eligible generators $(WindGenNonPTC)$ as well as current and future hourly demand within an ISO $(Load)$. Let $Load_{it}$ be the load (i.e., demand) in ISO $(i)$ for time $(t)$, where $t$ represents the clock time such that $t=0$ indicates the current hour relative to the dependent variable. Then, $t=1,\ldots,6$ allows future hourly load to control for ramping effects with $f(\cdot)$ being a flexible function form. Note that we can include hourly load on the right-hand side of our estimating equation because electricity demand is effectively exogenous at the hourly level \citep{hirth2024aggregate}, and, furthermore, we are not estimating a demand elasticity but employing load as a control only.

The full model may include the following fixed effects: $\omega_{my}$ and $\omega_{hm}$ denote year-by-month and month-by-hour fixed effects, respectively, while $\delta_{iy}$, $\delta_{im}$, and $\delta_{ih}$ signify ISO-by-year, -month, and -hour fixed effects, respectively. The year-by-month fixed effects creates an overall non-parametric time trend. The month-by-hour fixed effects account for differences in the average number of negative energy prices across months for the same hours and thus controls for seasonal effects. For example, 6am in February across the Eastern Interconnection is different than 6am in July. The ISO-year fixed effects generate ISO-specific, non-parametric time trends that augment the overall year-by-month time trend and account for changes in ISO composition such as total capacity and service area. The ISO-month and ISO-hour fixed effects adjust for any local factor particular to specific months or hours in the ISO. For example, it could be that local geography means the average wind potential is highest at 2am in a particular ISO. Finally, $\epsilon_{ihmy}$ is the error term, and errors are two-way clustered by $ISO$ and $Year \times Month$ for the pooled sample.\footnote{\citet{gu2019vcemway} note that a small-cluster correction factor is applied when clustering by ISO.}

The parameter $\beta$ is then causally identified in equation~\eqref{Eq:MainSpecification}, but due to the non-linear logit model, we report the average marginal effect (AME) in semi-log form to provide the change in the probability of an hour having a negative price as a function of percentage change in hourly PTC-eligible wind generation. This AME is a causal estimate too given the identifying assumptions outlined above. This causal effect includes the short-run effects of the extra wind energy production due to the subsidy as well as the aggregate investment decisions induced by the PTC. This allows us to understand that total effect of the PTC policy on negative prices. 

\vspace{2mm}
\noindent \textit{Other Mechanisms Specification.} We re-run equation~\eqref{Eq:MainSpecification} with a different dependent variable to check for the effects from other mechanisms on the probability of negative energy hour. Specifically, the new dependent variable, denoted $NegPriceAlt$, takes the value 1 when the energy price falls in the range $(-\infty,-PTC)$, and zero otherwise. In other words, this dependent variable indicates when a negative energy price should not be causally related to the PTC.

%%%%%%%%%%%%%%%%%%%%%%%%%%%%%%%%%%%%%%%%%%%%%%%%%%
%%%%%%%%%%%%%%%%%%%%%%%%%%%%%%%%%%%%%%%%%%%%%%%%%%
%%%%%%%%%%%%%%%%%%%%%%%%%%%%%%%%%%%%%%%%%%%%%%%%%%

\section{Results} \label{Sec:Results}

Table \ref{Tab:MainResults} provides our main results and reports the average marginal effect in semi-log form for our variable of interest $(WindGenPTC)$ as well as other important control variables. The columns show the results of separate regressions that vary in the set of controls and fixed effects following the specification in equation~\eqref{Eq:MainSpecification}. To start, column [1] is the simplest version of the estimating equation without fixed effects and only controls for the current ISO-level load. As predicted, increasing the amount of PTC-eligible wind generation increases the probability that an hour experiences a negative energy price. Specifically, increasing the PTC-eligible wind generation by 1 percent increases the probability of a negative energy price by 1.75 percentage points and is statistically significant (as indicated by the standard error in parentheses). We include the amount of non-PTC eligible wind generation and contemporaneous load -- denoted $WindGenNonPTC$ and $Load_0$, respectively -- in all specifications as controls. The control variable $WindGenNonPTC$ is not statistically significant in all specifications and thus indicates that the amount of non-PTC eligible wind generation has no effect on the probability of a negative price. That is, wind generation alone does not cause negative prices. Also, increasing the current period's load lowers the probability of a negative energy price hour, as predicted by our model, although not statistically significant in this basic specification.

%%%%%%%%%%%%%%%%%%%%%%%%%%%%%%%%%%%%%%%%%%%%%%%%%%
\begin{table}[ht]
    \centering
    \caption{Average Marginal Effect from Logistic Regression (Dep.\ Var.\ = $NegPrice$) }
    \resizebox{\columnwidth}{!}{%
    \begin{tabular}{rcccccccc}
    \hline \hline 
     Column: & [1] & [2] & [3] & [4] & [5] & [6] & [7] & [8] \\
    \hline
    \multicolumn{9}{c}{\textbf{Panel A: Semi-log Estimates in Percentage Points $(100 \times dy/ex)$}} \\
    \hline
    $WindGenPTC$    &  1.75  &  1.78  &  1.87  &  1.70  &  2.81  &  3.01  &  3.71  &  3.43   \\
                    & (0.60) & (0.62) & (0.69) & (0.67) & (0.71) & (0.58) & (0.70) & (0.92)  \\
    $WindGenNonPTC$ &  0.41  &  0.38  &  0.98  &  0.32  &  1.63  &  0.25  &  0.23  &  0.26   \\
                    & (0.71) & (0.66) & (0.72) & (0.63) & (1.07) & (0.47)& (0.54) & (0.78)  \\
    $Load_{0}$      & -3.56  &  -3.85 & -6.18  & -3.54  &-10.97  &-10.61  &-12.15  &-12.43   \\
                    & (1.94) & (1.66) & (2.53) & (1.46) & (1.13) & (2.19) & (1.95) & (1.68)  \\
    \hline
    \multicolumn{9}{c}{\textbf{Panel B: Controls and Fixed Effects}} \\
    \hline
    Future Load & & Yes & Yes & Yes & Yes & Yes & Yes & Yes \\             
    Year-Month $(\omega_{my})$ &  &  & Yes &     &     &     &     & Yes \\
    Month-Hour $(\omega_{mh})$ &  &  &     & Yes &     &     &     &     \\
    ISO-Year $(\delta_{iy})$   &  &  &     &     & Yes &     &     &     \\
    ISO-Month $(\delta_{im})$  &  &  &     &     &     & Yes &     & Yes \\
    ISO-Hour $(\delta_{ih})$   &  &  &     &     &     &     & Yes &     \\
    \hline
    \multicolumn{9}{c}{\textbf{Panel D: Summary Information}} \\
    \hline
    Observations $(N)$ & 419640 & 419352 & 398937 & 400138 & 375556 & 410426 & 322324 & 390755 \\
    Psudeo-$R^2$       & 0.222  & 0.239  & 0.292  & 0.261  & 0.346  & 0.345  & 0.300  & 0.371  \\
    \hline \hline
    \multicolumn{9}{p{15cm}}{\small{Notes: Each column reports results from a different regression. The standard errors reported in parenthesis are two-way clustered by $ISO$ and $Year \times Month$. Each semi-log estimate can be interpreted as the effect of a 1 percent change in the right-hand side variable on the probability of a negative price.  For instance, the coefficient on $WindGenPTC$ in column [1] is interpreted as a 1 percent increase in PTC eligible wind generation causes a 1.75 percentage point increase in the probability of a negative energy price.}}
    \end{tabular}}
    \label{Tab:MainResults}
\end{table}

%%%%%%%%%%%%%%%%%%%%%%%%%%%%%%%%%%%%%%%%%%%%%%%%%%

Column [2] reports results when including future ISO-specific load in the logistic regression to control for ramping effects that can contribute to negative prices (see Section \ref{Subsec:Ramping}). Specifically, we control for forward looking demand conditions using the first difference of load between the current and next period $(Load_{i1}-Load_{i0})$ as well as the difference between the fifth and sixth periods ahead $(Load_{i6}-Load_{i5})$. These two future load variables capture near- and long-term changes in demand, respectively, and are nearly uncorrelated. The orthogonality of these two variables avoids the multicollinearity that arises from including load in levels and the multicollinearity from including the entire vector of first differences. Here, the average marginal effects for $WindGenPTC$ and $WindGenNonPTC$ are virtually unchanged relative to column [1]. However, the magnitude of $Load_0$ increases to -3.85 percentage points and it is more accurately estimated.

Columns [3]-[8] add different combinations of fixed effects. Column [3] includes year-month fixed effects as a non-linear time trend and we find that average marginal effect for $WindGenPTC$ is slightly larger, but relatively unchanged compared to column [2], and the same pattern holds for the month-hour fixed effects in column [4]. However, the year-month fixed effects dramatically increase the magnitude of the $WindGenNonPTC$ and $Load_0$ average marginal effect estimates. Adding the ISO-level fixed effects across columns [5]-[8] qualitatively changes the results with a larger effect on the outcome of interest; for instance, column [6] employs ISO-month fixed effects and finds an average marginal effect of 3.01 percentage points for $WindGenPTC$. Column [6] also reports a much larger effect of contemporaneous load at -10.61 percentage points, while the effect from non-PTC eligible wind generators falls to 0.25 percentage points and is still not significant.

Our preferred specification in column [8] includes the year-month fixed effects as the non-linear time trend as well as ISO-month fixed effects to control for ISO-specific differences that vary across months.\footnote{The preferred specification does not include month-hour fixed effects as they do not significantly impact results, see column [4]. Also, ISO-year and ISO-hour fixed effects are not included as they dramatically decrease sample size. Regardless, the preferred specification employs 390,755 observations compared to the 419,640 observations in columns [2] since the fixed effects reduce the number of bins with variation in the dependent variable (including all fixed effects reduces the sample size to 272,061). Our main specification employs a logistic regression so the incidental parameters problem arises with many cross-sectional units and fixed effects \citep{wooldridge2010econometric}. This is an additional reason why we do not continue to stack fixed effects beyond the two sets included in our preferred specification in column [8].} 
This specification finds an average marginal effect semi-log estimate of 3.43 percentage points for the variable of interest and means increasing PTC-eligible wind generation by 1 percent then increases the probability of a negative energy price by 3.43 percentage points. This is a causal estimate (see Section~\ref{Sec:Empirical} for the identification assumptions). The point estimate in column [8] is nearly twice the estimate in column [3] indicating the importance of the ISO-level fixed effect in addition to the non-linear time trend when determining average marginal effect magnitudes.\footnote{We run a falsification test for the preferred specification by randomly assigning the share of wind generation that is PTC-eligible in each hour (i.e, permutation method). Then, we re-calculate the average marginal effect for $WindGenPTC$ from equation~\eqref{Eq:MainSpecification}. After 100 replications of random assignment, we conclude that the average marginal effect results passes the falsification test since the primary estimate from Table~\ref{Tab:MainResults} falls in the upper-5\% of the empirical CDF of the replication values with a p-value of less than 0.001.}

In addition to estimating the causal effect of the PTC on negative energy prices, we employ dominance analysis to determine how important the PTC is in explaining negative prices relative to other explanatory factors such as future load and the fixed effects \citep{luchman2021determining, bittmann2024primer}. When conducting the dominance analysis, we divide the explanatory variables into two groups and then calculate the share of the psudeo-$R^2$ statistic attributable to each group. The first group consists solely of the $WindGenPTC$ variable since it is our variable of interest. The second group contains all other explanatory variables including the fixed effects. For our preferred specification in Table \ref{Tab:MainResults} column [8], we find that 9.99 percent of the variance is explained by $WindGenPTC$ alone. Thus, approximately 10 times as much of the variance is explained by all other factors with respect to negative prices. This means that the PTC subsidy has a qualitatively important but relatively small effect in this pooled sample when explaining the prevalence of negative prices.

Next, we consider whether the PTC has heterogeneous treatment effects across ISOs. Table~\ref{Tab:HeterogeneousResults} reports estimates for $WindGenPTC$ where each column is a separate logistic regression by ISO. The estimating equation used for each column is similar to our preferred pooled specification, but the ISO-specific fixed effects no longer apply.\footnote{The total observations across columns in Table \ref{Tab:HeterogeneousResults} is less than 390,755 from the preferred pooled specification because additional $Year \times Month$ bins have no variation in the dependent variable when the sample is split by ISO.}
The average marginal effect estimates for $WindGenPTC$ appear in Panel A while the summary information for each regression can be found in Panel B. Table~\ref{Tab:HeterogeneousResults} finds that the point estimates for ISO-NE and PJM are near zero, MISO and NYISO have estimates nearer to the pooled estimate, while the estimate for SPP is larger by an order of magnitude. The SPP point estimate implies that a 1 percent increase in PTC-eligible wind generation increases the probability of a negative energy price hour by 25.0 percentage points. This large point estimate must be put in the context that only 4 percent of SPP's observed prices fall in the range $[-PTC,0)$. Therefore, the large point estimate for SPP only implies that the probability of a given hour becoming negative increases from 4 to 5 percent.  Interestingly, NYISO has a higher raw probability of negative energy prices than SPP, but a much lower point estimate for $WindGenPTC$ than SPP. 

Table \ref{Tab:HeterogeneousResults} Panel C reports the dominance analysis for each ISO regression and also finds significant heterogeneity. As in the pooled sample dominance analysis, $WindGenPTC$ is in one group and all other explanatory variables are in the other group. In PJM, the $WindGenPTC$ variable only explains 3.6 percent of the total variance in the model, but for SPP the same variable explains 29.0 percent. Meanwhile, ISO-NE and MISO have similar dominance analysis shares for $WindGenPTC$ but significantly different point estimates.

%%%%%%%%%%%%%%%%%%%%%%%%%%%%%%%%%%%%%%%%%%%%%%%%%%

\begin{table}[t]
    \centering
    \caption{Average Marginal Effect Heterogeneous Estimates by ISO (Dep.\ Var.\ = $NegPrice$)}
    \begin{tabular}{rccccc}
    \hline \hline 
     Column:     & [1]   & [2]   & [3]  & [4] & [5] \\
     ISO/RTO: & ISONE & MISO & NYISO & PJM & SPP \\
    \hline
    \multicolumn{6}{c}{\textbf{Semi-log Estimates in Percentage Points} $(100 \times dy/ex)$} \\
    \hline
    $WindGenPTC$    &  0.46  &  1.58  &  4.83  &  0.39  &  25.04  \\
                    & (0.23) & (0.18) & (0.57) & (0.11) & (1.79) \\
    \hline
    \multicolumn{6}{c}{\textbf{Panel B: Summary Information}} \\
    \hline
    Pr($NegPrice=1$) [in \%]     & 0.495  & 2.98  & 4.68  & 0.11  & 4.14 \\
    Observations $(N)$        & 34252  & 51866  & 55938  & 29297  & 77185 \\
    Psudeo-$R^2$              & 0.270  & 0.365  & 0.260  & 0.328  & 0.381   \\
    \hline 
    \multicolumn{6}{c}{\textbf{Panel C: Dominance Analysis in Percentage Points}} \\
    \hline
    $WindGenPTC$                       & 3.92  &  4.22  &  14.55  &  3.62  &  29.00  \\
    \hline \hline
    \multicolumn{6}{p{11cm}}{\small{
    Notes: Each column reports results from a different regression using ISO-specific observations. The standard errors in parenthesis are clustered by $Year \times Month$. Each semi-log estimate can be interpreted as the effect of a 1 percent change in the right-hand side variable on the probability of a negative price.  For instance, the coefficient on $WindGenPTC$ in column [1] is interpreted as a 1 percent increase in PTC eligible wind generation causes a 0.46 percentage point increase in the probability of a negative energy price.}}
    \end{tabular}
    \label{Tab:HeterogeneousResults}
\end{table}

%%%%%%%%%%%%%%%%%%%%%%%%%%%%%%%%%%%%%%%%%%%%%%%%%%

\begin{table}[htb]
    \centering
    \caption{Average Marginal Effect from Logistic Regression (Dep.\ Var.\ = $NegPriceAlt$)}
    \begin{tabular}{rcccccccc}
    \hline \hline 
     Column: & [1] & [2] & [3] & [4] & [5] & [6] & [7] & [8] \\
    \hline
    \multicolumn{9}{c}{\textbf{\textbf{Semi-log Estimates in Percentage Points} $(100 \times dy/ex)$}} \\
    \hline
    $WindGenPTC$    &  0.56  &  0.57  &  0.55  &  0.67  &  0.84  &  0.67  &  0.87  &  0.94  \\
                    & (0.07) & (0.08) & (0.14) & (0.13) & (0.17) & (0.19) & (0.28) & (0.10) \\
    $WindGenNonPTC$ & -0.21  & -0.21  &  0.01  & -0.25  & -0.16  & -0.19  & -0.25  & -0.20  \\
                    & (0.01) & (0.01) & (0.10) & (0.06) & (0.20) & (0.08) & (0.11) & (0.06) \\
    $Load_{0}$      & -0.39  &  -0.50 & -0.98  & -0.54  & -2.46  & -1.92  & -2.41  & -2.65 \\
                    & (0.12) & (0.13) & (0.48) & (0.15) & (0.62) & (0.68) & (0.64) & (0.80) \\
    \hline
    \multicolumn{9}{c}{\textbf{Panel B: Controls and Fixed Effects}} \\
    \hline
    Future Load                & & Yes & Yes & Yes & Yes & Yes & Yes & Yes \\          
    Year-Month $(\omega_{my})$ &  &  & Yes &     &     &     &     & Yes\\
    Month-Hour $(\omega_{mh})$ &  &  &     & Yes &     &     &     &\\
    ISO-Year $(\delta_{iy})$   &  &  &     &     & Yes &     &     &\\
    ISO-Month $(\delta_{im})$  &  &  &     &     &     & Yes &     & Yes\\
    ISO-Hour $(\delta_{ih})$   &  &  &     &     &     &     & Yes & \\
    \hline
    \multicolumn{9}{c}{\textbf{Panel C: Summary Information}} \\
    \hline
    Observations $(N)$ & 419640 & 419352 & 371103 & 319014 & 314161 & 382333 & 266920 & 339172 \\
    Psudeo-$R^2$       & 0.141  & 0.166  & 0.237  & 0.188  & 0.222  & 0.231  & 0.195  & 0.280 \\
    \hline \hline
    \multicolumn{9}{p{15cm}}{\small{   
    Notes: Each column reports results from a different regression. The standard errors reported in parenthesis are two-way clustered by $ISO$ and $Year \times Month$. Each semi-log estimate can be interpreted as the effect of a 1 percent change in the right-hand side variable on the probability of a negative price.  For instance, the coefficient on $WindGenPTC$ in column [1] is interpreted as a 1 percent increase in PTC eligible wind generation causes a 0.56 percentage point increase in the probability of a negative energy price.}}
    \end{tabular}
    \label{Tab:AltResults}
\end{table}

%%%%%%%%%%%%%%%%%%%%%%%%%%%%%%%%%%%%%%%%%%%%%%%%%%

Next, we turn our attention to examining hours where the PTC alone could \textit{not} be responsible for negative prices.  Table \ref{Tab:AltResults} reports results for the hourly regression alternative mechanism specification (see Section~\ref{Sec:Empirical} for details). Again, Column [1] is the simplest version of the estimation equation without fixed effects and only controls for the contemporaneous ISO-level load. For this specification, the estimate for the average marginal effect for $WindGenPTC$ is 0.56 percentage points and therefore significantly smaller than the corresponding estimate in Table~\ref{Tab:MainResults}. This suggests that other production subsidies, such as renewable energy credits (RECs) play a role in the market. This small point estimate doubles for the preferred specification in column [8], but is still significantly smaller than in the main results table. We also note that the $WindGenNonPTC$ estimates often have the wrong sign (i.e., negative) relative to expectations, and the $Load_0$ estimate is much smaller too. However, if we move the cutoff to 1.5 times the PTC value, that is, $NegPriceAlt$ equal 1 if the energy price is in the range $(-\infty,1.5 \times -PTC)$, then the average marginal effect estimate of the PTC for the preferred specification falls to only 0.2 percentage points and is no longer statistically significant, as predicted by our model. Moving this threshold plausibly reduces impacts from other subsidies that PTC-eligible generators might also receive from state or local authorities.

%%%%%%%%%%%%%%%%%%%%%%%%%%%%%%%%%%%%%%%%%%%%%%%%%%
%%%%%%%%%%%%%%%%%%%%%%%%%%%%%%%%%%%%%%%%%%%%%%%%%%
%%%%%%%%%%%%%%%%%%%%%%%%%%%%%%%%%%%%%%%%%%%%%%%%%%

\section{Discussion} \label{Sec:Welfare}

Our empirical results disentangle the effect of the PTC on the frequency of negative prices from other causes as well as bound the pecuniary costs of the PTC to electricity generators. This allows us to estimate the number of hours that have negative energy prices due to the PTC over our sample period.  Additionally, our model demonstrates that there is a minimum negative price that any production subsidy can induce a negative price: the value of the production subsidy.  Therefore, we are able to combine our regression estimates with the time-varying value of the PTC to estimate the potential costs of the PTC to electricity generators.

For our preferred regression specification from Table \ref{Tab:MainResults}, we conducted a dominance analysis to infer the contribution of the PTC to the probability of observing an hour with a negative price.  We found that the PTC explains just 9.99 percent of the probability of an hour having a negative price. Using the dominance analysis result, we compute an upper bound on expenditures due to the PTC.  To do this, we assume that the 9.99 percent of hours with a negative price closest to the value of the PTC have negative prices due to the PTC while other hours with negative prices are caused by physical ramping constraints (or other mechanisms).  These calculations suggest that a maximum of \$369 million in expenditures out of a total of \$2.8 billion in expenditures from negative prices could be due to the PTC during the 2014-2022 period.\footnote{We examine expenditures by electricity generators and benefits over the 2014-2022 period since we have all necessary data from all ISOs during this period.  This facilitates a meaningful comparison across ISOs.}  Thus, if the PTC had not been in place over our sample period, expenditures by generators would be 13 percent lower than we observe in the pooled sample.  These results are shown in the left column of Table \ref{Tab:MissMkt} in Panel A.

\begin{table}[ht]
  \centering
  \caption{Decomposition of Expenditures (nominal dollars, millions)}
    \begin{tabular}{lrrrrrrr}
    \hline\hline
    \multicolumn{8}{c}{\textbf{Panel A: Negative-Price Expenditures}} \\
    \hline
    && \multicolumn{6}{c}{Heterogeneous Effects by ISO} \\ \cline{3-8}
          & Pooled & ISO-NE & MISO & NYISO & PJM & SPP & Total \\
    \hline
    All Expenditures & 2,823 & 221   & 25    & 1,176 & 137   & 1,265 & 2,823 \\
    PTC Induced Expenditures & 369   & 4     & 3     & 97    & 7     & 459   & 570 \\
    Expenditures without PTC & 2,453 & 217   & 21    & 1,079 & 130   & 806   & 2,252 \\
    Expenditures without PTC $(\%\Delta)$ & -13\% & -2\%  & -14\% & -8\%  & -5\%  & -36\% & -20\% \\
    \hline
    \multicolumn{8}{c}{\textbf{Panel B: Benefits from PTC}} \\
    \hline
    Environmental Benefits & 655   & $>$1   & 5     & 11    & 1     & 1,075 & 1,092 \\
    Net Benefits & 286   & -4    & 1     & 86   & -6    & 617   & 521 \\
    \hline\hline
\multicolumn{8}{p{5.75in}}{\small{Notes: All values are calculated over 2014-2022, since we have complete data for all ISOs in this period to enable meaningful comparisons across ISOs. The estimated environmental benefits for ISO-NE is \$0.3 million.}}
    \end{tabular} \\
  \label{Tab:MissMkt}
\end{table}

Since we estimate heterogeneous effects of the PTC across electricity markets as Table~\ref{Tab:HeterogeneousResults} reports, we separately calculate the effect of the PTC on energy expenditures by generators for each market.  These are shown in the remaining columns of Panel A in Table \ref{Tab:MissMkt} using the dominance analysis reported in Panel~C of Table~\ref{Tab:HeterogeneousResults}.  Given these ISO-specific estimates, we estimate an upper bound of expenditures that is 54 percent larger than the pooled sample; specifically, \$570 million and most of those expenditures occur in SPP. Also, in SPP, the PTC related expenditures account for 36 percent of total negative price expenditures by electricity generators, by far the largest in percentage terms. Meanwhile, in NYISO, electricity expenditures would have decreased by only 8 percent and this is due in part to the large number of negative prices hours below $-PTC$ in that ISO (see Figure~\ref{fig:PriceDistribution}).

Up to this point, we have exclusively examined the costs to electricity generators of the PTC.  However, the PTC was enacted with an implicit objective of reducing local and global pollution from electricity generators.  Therefore, to better understand the PTC's costs, we compare the costs to benefits accrued due to the lower levels of local and global pollution in the hours in which we observe negative prices. Then, we compare PTC related energy expenditures to the estimated environmental benefits from wind generation that is eligible for the PTC \textit{in that hour}.  As far as the PTC induced additional wind capacity, the benefits of the PTC are accrued across all hours that those wind turbines generate electricity, but our comparison leaves out the marginal benefits from the wind generation in hours where prices are not negative.

To calculate the environmental benefits of additional wind generation in any hour, we use estimates from \cite{Holland2016}.  They estimate the marginal environmental damages of air pollution from an additional megawatt of electricity generation that vary by hour and ISO.  These estimates include damages from CO2, SO2, NOx, PM2.5, and VOCs.  We assume that each additional megawatt-hour of wind generation leads to an equivalent emissions reduction from the marginal source, so that \citet{Holland2016}'s damage estimates are equivalent to the environmental benefit of an additional megawatt of wind.\footnote{\cite{Holland2016} estimates are based on a \$6 million (2000\$) value of a statistical life.}  

To compute the counterfactual emissions, in every hour that has a negative price caused by the PTC, we calculate the total PTC eligible wind generation in that ISO and multiply the ISO generation by the marginal benefits from wind for the ISO for each megawatt hour.  We then sum those benefits across our sample to estimate the total environmental benefits of wind generation.  These environmental benefits are displayed in Table \ref{Tab:MissMkt} Panel B. The benefits from these avoided emissions range from \$655-1,092 million.

The bottom row of Panel B subtracts the PTC-induced expenditures in Panel A from the environmental benefits in to top row of Panel B to yield the net benefits of the additional wind generation.  This row shows that the PTC creates net benefits across our sample between \$286 million and \$521 million in negative priced hours due to the PTC. Notably, these benefits are concentrated in SPP with all other ISOs showing a relatively small net loss (and a small net gain in MISO) from the PTC in negative price hours.  While SPP is estimated to have the largest percentage of expenditures induced by the PTC, the net benefits are also the largest in SPP, leading to an estimated \$617 million in net benefits over our sample.  NYISO experiences the largest net loss of \$86 million. In other words, the cost of the PTC in the form of energy expenditures exceeds the environmental benefits \textit{in those hours} when prices are negative.  However, there are also environmental benefits in hours in which PTC eligible generators produce electricity with positive electricity prices that we are not capturing in this calculation, and the environmental benefits in positive price hours are larger over the course of the generators operational life than the costs of the negative electricity prices.

%%%%%%%%%%%%%%%%%%%%%%%%%%%%%%%%%%%%%%%%%%%%%%%%%%
%%%%%%%%%%%%%%%%%%%%%%%%%%%%%%%%%%%%%%%%%%%%%%%%%%
%%%%%%%%%%%%%%%%%%%%%%%%%%%%%%%%%%%%%%%%%%%%%%%%%%

\section{Conclusion} \label{Sec:Conclusion}

This study estimates the causal effect of the U.S. federal production tax credit for renewable generation (i.e., wind) on the probability of negative energy prices in wholesale electricity markets. The preferred point estimate reveals that a one percent increase in PTC-eligible wind generation leads to a 3.4 percentage point increase in the probability of a negative energy price in a given hour. We also find that only 10 percent of the variance explaining the prevalence of negative prices is attributable to PTC-eligible wind generation. This suggests that while the PTC contributes to negative prices, it is not often the primary factor.  However, significant heterogeneity exists across electricity markets, as the PTC is much more likely to cause negative energy prices in SPP than in the other ISOs examined in this study. 

Using our causal estimate, we calculate the amount generators paid to produce electricity during the period 2014-2022.  We find that an upper-bound of payments due to the PTC falls between \$369 and \$570 million over this 9-year period compared to total payments from negative prices of \$2.8 billion. We next compare these costs to estimated environmental benefits from additional wind generation in the hours with negative prices caused by the PTC, and conclude that the benefits are at least twice the costs to generators.  However, these benefits are highly concentrated in the SPP market while the net benefits are close to zero in the other markets. 

Our work contributes to understanding the effects of increasing the share of renewable electricity generation on electricity market pricing. It also contributes to understanding of the causes of negative energy prices and suggests that production subsidies are not the primary cause of negative prices in most electricity markets. Consequently, the expansion of electricity storage and transmission capacity will likely have a greater effect on decreasing the prevalence of negative prices than the phase-out of production subsides. 

%%%%%%%%%%%%%%%%%%%%%%%%%%%%%%%%%%%%%%%%%%%%%%%%%%
%%%%%%%%%%%%%%%%%%%%%%%%%%%%%%%%%%%%%%%%%%%%%%%%%%
%%%%%%%%%%%%%%%%%%%%%%%%%%%%%%%%%%%%%%%%%%%%%%%%%%

\bibliography{np.bib}
\bibstyle{aer}

%%%%%%%%%%%%%%%%%%%%%%%%%%%%%%%%%%%%%%%%%%%%%%%%%%
%%%%%%%%%%%%%%%%%%%%%%%%%%%%%%%%%%%%%%%%%%%%%%%%%%
%%%%%%%%%%%%%%%%%%%%%%%%%%%%%%%%%%%%%%%%%%%%%%%%%%

\end{document}